\documentclass[a4paper]{JHEP3}
\newtheorem{theorem}{No-Go Theorem}
\title{Taming the Leibniz Rule on the Lattice}
\author{Mitsuhiro Kato\\
        Institute of Physics, University of Tokyo, Komaba, 
        Meguro-ku, Tokyo 153-8902, Japan\\
        E-mail: \email{kato@hep1.c.u-tokyo.ac.jp}
        }
\author{Makoto Sakamoto\\
        Department of Physics, Kobe University, Nada-ku, 
        Hyogo  657-8501, Japan\\
        E-mail: \email{dragon@kobe-u.ac.jp}
        }
\author{Hiroto So\\
        Department of Physics,  Ehime University, Bunkyou-chou 2-5, 
        Matsuyama 790-8577, Japan\\
        E-mail: \email{so@phys.sci.ehime-u.ac.jp}
        }
\abstract{
We study a product rule and a difference operator equipped
with Leibniz rule in a general framework of lattice field theory.
It is shown that the difference operator can be determined by 
the product rule and some initial data through the Leibniz rule.
This observation leads to a no-go theorem that it is impossible
to construct any difference operator and product rule on a
lattice with the properties of (i) translation invariance,
(ii) locality and (iii) Leibniz rule.
We present a formalism to overcome the difficulty by an infinite 
flavor extension or a matrix expression of a lattice field theory.
}
\preprint{UT-Komaba/08-5, KOBE-TH-08-03, Ehime-th-6}
\begin{document}
%
%
%

%
%
%
\section{Introduction}
%
%
%

It is of great importance to formulate supersymmetric theories
on a lattice to study non-perturbative dynamics, especially
supersymmetry breaking.
Over the last thirty years, a considerable number of attempts
have been made to construct lattice supersymmetric models
\cite{Feo2003,Kaplan2004,Giedt}.
However, none of them have not fully succeeded in realizing
supersymmetry on a lattice.
A key to construct interacting supersymmetric theories is
to keep a Leibniz rule on a lattice \cite{Dondi-Nicolai,
Fujikawa}.\footnote{
Recently, novel ideas of the noncommutativity approach
\cite{Kawamoto2005} and the link approach \cite{Kawamoto2006}
have been proposed to restore a Leibniz rule for
supersymmetry transformations in twisted supersymmetric models
on the lattice.
Further investigation, however, seems to be necessary
\cite{Bruckmann2006,Bruckmann2007}.
}
Naive difference operators, like forward/backward/symmetric
ones, do not satisfy a Leibniz rule.
Lattice models equipped with a Leibniz rule exist by allowing
the non-locality of interactions
or difference operators \cite{Dondi-Nicolai, SLAC, 
Bartels-Bronzan, Nojiri, Bouguenaya-Fairlie}.
In fact, it is pointed out that it is difficult to impose
simultaneously the following three properties:
(i) translation invariance, (ii) locality and (iii) Leibniz rule
in any lattice field theories \cite{So-Ukita}.\footnote{
A no-go theorem in a restricted case was given 
in Ref.\cite{SLAC, Bouguenaya-Fairlie}.
}

In this article, we prove the above statement as a no-go 
theorem in general lattice theories. 
The requirement of the associative law leads us to an easier
proof of the no-go theorem, but it is not necessary to 
the proof.
We further show that it is impossible to solve the Leibniz
rule problem even if a product rule of fields and a 
difference operator are extended to include multi-flavor
indices.
Our proof shows that a difference operator can be determined
from information of a product rule and some initial data
through the equation derived from a Leibniz rule.
Then, it turns out that any local product rule inevitably
leads to a non-local difference operator in any translationally
invariant lattice theories of finite flavors.
One way to escape from the no-go theorem is to introduce
an infinite number of flavors and a nontrivial connection
between lattice sites and flavors.
We propose a translationally invariant local lattice theory
that a difference operator satisfying a Leibniz rule is
realized with a product rule equipped with an associative
law in a matrix formulation.

In section 2, our fundamental tools of product rule,
difference operator, translational invariance and locality
are explained.
In section 3, we see that the associative law restricts
the form of the product rule essentially to a normal 
local product.
We prove the no-go theorem for general one-flavor systems
in section 4 and for general multi-flavor systems in section 5.
In section 6, we present a lattice model that evades the
no-go theorem by introducing an infinite number of flavors.
Section 7 is devoted to summary and discussions.

\vspace{5mm}
%
%
%
\section{Locality of product rule and difference operator 
on a lattice}
%
%
%

A lattice gauge theory has usually treated only 
ultra local operators except Dirac operators  
such as Ginsparg-Wilson fermion or overlap-Dirac operator
\cite{Ginsparg-Wilson,Neuberger,HJL}. 
In order to analyze  a Leibniz rule and an associative law 
on a lattice, we must generalize a product rule between fields 
and a difference operator on a field. 

A lattice field product between $\phi_n$  and $\psi_n$
is defined as
%
\begin{equation}
(\phi \cdot \psi)_n \equiv \sum_{l,m} C_{lmn}\phi_l \psi_m\,  ,
\label{product}
\end{equation} 
%
where the coefficient $C_{lmn}$ becomes  a key of this 
product definition keeping bi-linearity on  both fields. 
The indices $l,m,n$ imply 
positions on a lattice which has  an infinite size.
Although we restrict our consideration to one-dimensional
lattice throughout this paper, the extension to higher 
dimensions will be straightforward.
If one chooses  
%
\begin{equation}
C_{lmn} = \delta_{l,n}\delta_{m,n}\,, 
\label{eq2}
\end{equation}
%
as the product rule,  
then it defines 
the normal product of lattice fields at the same point.

Another coefficient $D$ on a field
%
\begin{equation}
(D\phi)_n \equiv \sum_m D_{mn}\phi_m
\label{diff}
\end{equation}
%
means a generalized difference operator keeping the linearity 
about the field. 
The difference operator for a constant field implies 
%
\begin{equation}
\sum_m D_{mn} = 0\,.  
\label{eq4}
\end{equation}
%
Two familiar examples for $D_{mn}$ are the forward and backward 
difference operators defined, respectively, as
%
\begin{eqnarray}
\label{eq5}
D_{mn}^+ &=& \delta_{m,n+1} -\delta_{m,n}\,,\\
\label{eq6}
D_{mn}^- &=& \delta_{m,n} - \delta_{m,n-1}\, .
\end{eqnarray}
%

If the system has no external field, it should keep 
translational invariance. 
The invariance for $C_{lmn}$ and $D_{mn}$ is imposed 
as the following forms:
%
\begin{eqnarray}
\label{eq7}
C_{lmn} &=& C(l-n,m-n)\,, \\
\label{eq8}
D_{mn} &=& D(m-n)\,. 
\end{eqnarray}
%

The locality property is important in constructing local 
field theories after the continuum limit. 
To make the locality manifest, we define Fourier transform
of the coefficients $C(k,l)$ and $D(m)$ by
%
\begin{eqnarray}
\label{hatC}
\hat{C}(v,w) &\equiv&  \sum_{k,l=-\infty}^{\infty}C(k,l)v^kw^l ,\\
\label{hatD}
\hat{D}(z) &\equiv&  \sum_{m=-\infty}^{\infty}D(m)z^m ,
\end{eqnarray}
where $v, w, z$ are $S^{1}$-variables given by
$v=e^{ip},~w=e^{iq},~z=e^{ir}$ and will be extended to
some complex domains later.
As we will explain below, the locality of the product rule
(\ref{product}) and the difference operator (\ref{diff})
is directly related to the holomorphic property of the
complex functions.

In terms of the complex function $\hat{D}(z)$, 
the condition (\ref{eq4}) can be rewritten as 
%
\begin{equation}
\hat{D}(1)=0\,, 
\label{init-cond}
\end{equation}
%
which may be regarded as an initial condition of the
function $\hat{D}(z)$.

From our knowledge about lattice fields and complex  analysis, 
the local property of the product rule and the difference 
operator restricts us to holomorphic functions for 
$\hat{C}(v,w)$ and $\hat{D}(z)$.  
To discuss more strictly,  we prepare an annulus 
${\cal D}_2=\{(v,w) | 1-\epsilon <|v|, |w| <1+  \epsilon  \} $
for  $\hat{C}(v,w)$ and another annulus 
${\cal D}_1= \{z| 1- \epsilon < |z| < 1+  \epsilon \}$
for $\hat{D}(z)$, where $\epsilon$ is a positive constant
smaller than unity. 
The functions $\hat{C}(v,w)$ and $\hat{D}(z)$ are 
analytically extended to these annulus domains uniquely 
owing to their holomorphism. 
We state a lemma about the locality of $C(k,l)$ here.

\vspace{4mm}
\noindent
{\bf Lemma}\ \ 
{\it 
The following  two propositions are equivalent to each other:
%
\begin{itemize}
\item[1.]A product rule $C(k,l)$ is local. 
\vspace{-3mm}
\item[2.]The corresponding $\hat{C}(v,w)$ is holomorphic
on ${\cal D}_2$.
\end{itemize}
%
}
%
%
\vspace{2mm}

\noindent
If the proposition 1 holds, $C(k,l)$ is exponentially decaying 
as $o(\exp (-r_1|k|), \exp (-r_2 |l|) )$ for large $|k|$ and 
$|l|$ where $r_1$ and $r_2 $ are some positive numbers. 
From this behavior of $C(k,l)$,  we can define a complex 
function $\hat{C}(v,w)$ as
%
\begin{equation}
\hat{C}(v,w)  \equiv \sum_{k,l}C(k,l)v^kw^l ,
\label{define-of-C}
\end{equation}
%
which is uniformly convergent in 
$\{(v,w)|e^{-r_1}<|v|< e^{r_1}, e^{-r_2}<|w|< e^{r_2} \}$ 
and is holomorphic on ${\cal D}_2$ where $1-\epsilon = e^{-r_{1}}$
for $r_{1} < r_{2}$ ($1-\epsilon = e^{-r_{2}}$ for $r_{2} < r_{1}$).  
Conversely, if the proposition 2 holds,  eq.(\ref{define-of-C}) 
shows the Laurent expansion with $v$ and $w$, which
converges on the annulus ${\cal D}_2$. 
The coefficient $C(k,l)$ for positive integers $k,l$ 
behaves, with $0 < \epsilon' < \epsilon$, as 
%
\begin{equation}
|C(k,l)|   = \bigg|\oint_{|v|= 1+\epsilon'} \frac{dv}{2\pi i}
\oint_{|w|= 1+\epsilon'} \frac{dw}{2\pi i} 
\hat{C}(v,w)v^{-k-1}w^{-l-1}   \bigg|
\le   K_1(1+\epsilon')^{-k-l}  ,
\label{radius_1}
\end{equation}
%
for negative integers $k,l$, 
%
\begin{equation}
|C(k,l)| =  \bigg|\oint_{|v|= 1-\epsilon'}  \frac{dv}{2\pi i}
 \oint_{|w|= 1-\epsilon'} \frac{dw}{2\pi i}  
 \hat{C}(v,w)v^{-k-1} w^{-l-1}  \bigg|
 \le  K_2(1 -\epsilon')^{-k-l} , 
\label{radius_2}
\end{equation} 
%
for a positive $k$ and a negative $l$, 
%
\begin{equation}
|C(k,l)| =  \bigg|\oint_{|v|= 1+\epsilon'} \frac{dv}{2\pi i}
 \oint_{|w|= 1-\epsilon'} \frac{dw}{2\pi i}  
 \hat{C}(v,w)v^{-k-1}w^{-l-1}  \bigg|
 \le  K_3(1+\epsilon')^{-k}(1 -\epsilon')^{-l} , 
\label{radius_3}
\end{equation} 
%
and a negative $k$ and a positive $l$, 
%
\begin{equation}
 |C(k,l)| =  \bigg|\oint_{|v|= 1-\epsilon'} \frac{dv}{2\pi i} 
 \oint_{|w|= 1+\epsilon'}\frac{dw}{2\pi i} 
 \hat{C}(v,w)v^{-k-1}w^{-l-1}   \bigg|
 \le  K_4(1-\epsilon')^{-k}(1 +\epsilon')^{-l} , 
\label{radius_4}
\end{equation} 
%
where $K_1,K_2,K_3$ and $K_4$  are finite and positive constants 
because the absolute value of the holomorphic function 
$\hat{C}(v,w)$ is finite on ${\cal D}_2$.   
The relations (\ref{radius_1})$\sim$(\ref{radius_4}) imply 
that $C(k,l)$ is decaying with  $|k|$ and $|l|$ exponentially. 
Therefore,  $C(k,l)$ is local.\footnote{
If we impose the smoothness on $\hat{C}(v,w)$ instead of 
the holomorphy,  $C(l,k)$ is permitted to behave as power-damping.
}
In a similar way, we can show that if $D(m)$ is local,
then the corresponding complex function $\hat{D}(z)$ is
holomorphic on an annulus ${\cal D}_1$, and vice versa.

It is meaningful to define a terminology, \lq\lq a trivial product 
rule\rq\rq, ${\hat C}(v,w)$ which is identically zero on the 
defined domain ${\cal D}_2$. 
Then, any field on a lattice multiplied by another field 
becomes vanishing and it is impossible to construct 
nontrivial theories.    
We can always find a 2-dimensional complex subdomain
${\cal F}_2 = \{ (v,w)|\ \hat{C}(v,w) \ne 0 \}$ in  ${\cal D}_2$ 
in considering a nontrivial product rule ${\hat C}(v,w)$, 
otherwise the function ${\hat C}(v,w)$ is identically zero 
because of the identity theorem on complex functions.

\vspace{5mm}
%
%
%
\section{Associative law and product rule on a lattice}
%
%
%
In cases of interacting theories, we must consider field 
products of three-body or more. 
The consistency of field products in an actual model is
often controlled by additional requirements like associativity.
In this section, we examine the product rule that 
satisfies the associative law   
$(\phi \cdot \psi)\cdot \chi = \phi \cdot (\psi\cdot \chi)$. 
This can be read as
%
\begin{equation}
\label{eq17}
\sum_jC_{lmj}C_{jnk}  = \sum_jC_{ljk}C_{mnj}\,. 
\end{equation}
%
After the translational invariance is imposed, 
the law can be rewritten, by use of the holomorphic 
function (\ref{define-of-C}), as 
%
\begin{equation}
\hat{C}(v,w)\hat{C}(vw,z) = \hat{C}(v,wz)\hat{C}(w,z)\,.
\label{hol-associative-law}
\end{equation}
%
From eq.(\ref{hol-associative-law}) and the holomorphy, it turns out that
any nontrivial product rule $\hat{C}(v,w)$ can always be
expressed, in terms of a holomorphic function $F(v)$ on
${\cal D}_1$, as
%
\begin{equation}
\hat{C}(v,w) = \frac{F(vw)}{F(v)F(w)}\, .
\label{sol-assoc}
\end{equation}
%
The proof is given in Appendix.

To investigate the meaning of a factorization in eq.(\ref{sol-assoc}), 
we redefine a local field on a lattice as 
%
\begin{equation}
\phi_n = \sum_{m}a_{mn}\phi^{\prime}_{m}\, ,
\label{eq20}
\end{equation}
%
where the translational invariance and the locality are 
imposed, i.e. $a_{mn}= a(m-n)$ and $\hat{a}(v) = \sum_{m} a(m)v^m$. 
After the local redefinition (\ref{eq20}) of fields, 
our product rule is transformed as 
%
\begin{equation}
\hat{C}'(v,w) = \frac{\hat{a}(v)\hat{a}(w)}{\hat{a}(vw)}\hat{C}(v,w)\,.
\label{eq21}
\end{equation}
%
This implies that we can always set $\hat{C}'(v,w)=1$, 
by choosing $\hat{a}(v)=F(v)$, which is nothing but the
normal local product $C'_{lmn} = \delta_{l,n} \delta_{m,n}$.
Therefore, we conclude that the product rule satisfying the
associative law (\ref{eq17}) or (\ref{hol-associative-law}) 
is essentially unique and is given by the normal product (\ref{eq2}).

\vspace{5mm}
%
%
%
\section{No-go theorem}
%
%
%
In this section, we prove no-go theorems about a Leibniz rule
on a lattice. 
We first assume the associative law for a product rule but 
later we give a proof without referring to the condition.  
The statement of the no-go theorem we first present is given 
as follows:
%

%
%
\begin{theorem}
\label{theorem1}
It is impossible to construct a lattice field theory in an infinite
lattice volume with a nontrivial product rule (\ref{product})
and a difference operator (\ref{diff}) that satisfy the
following four properties:
(i) translation invariance, (ii) locality,  (iii) Leibniz rule 
and (iv) associative law.
\end{theorem}
%
%

\noindent
The proof is simple and goes as follows.
A Leibniz rule
$D(\phi\cdot\psi) = (D\phi)\cdot\psi + \phi\cdot(D\psi)$
can be translated into a relation between the product rule
and the difference operator as
%
\begin{equation}
\sum_{k} C_{lmk} D_{kn}  = \sum_{k} C_{kmn} D_{lk}  
  + \sum_{k} C_{lkn} D_{mk}\,. 
\label{leibniz}
\end{equation}
%
With the properties (i) and (ii), the relation (\ref{leibniz})
can be rewritten, in terms of the holomorphic functions
$\hat{C}(v,w)$ and $\hat{D}(z)$ defined in eqs.(\ref{hatC}) 
and (\ref{hatD}), as
%
\begin{equation}
\hat{C}(v,w)\Big(\hat{D}(vw) - \hat{D}(v)- \hat{D}(w)\Big) = 0\, . 
\label{leibniz4}
\end{equation}
%
Since  $\hat{C}(v,w)$ satisfying the associative law can be 
set to unity, as shown in the previous section, 
the condition (\ref{leibniz4}) reduces to
%
\begin{equation}
\hat{D}(vw) - \hat{D}(v) - \hat{D}(w) = 0\,.
\label{leibniz5}
\end{equation}
%
Differentiating it with respect to $v$ and then putting
$v=1$, we have
%
\begin{equation}
w\partial_{w}\hat{D}(w) = \partial_{v} \hat{D}(v)\big|_{v=1}\,.
\label{diff-for-L}
\end{equation}
%
With the initial condition (\ref{init-cond}), the solution 
to eq.(\ref{diff-for-L}) is given by
%
\begin{equation}
 \hat{D}(w) =  \beta \log w\,, 
\label{log-sol}
\end{equation}
%
where $\beta = \partial_{v} \hat{D}(v)|_{v=1}$.
The coefficient $\beta$, however, has to vanish, otherwise
the difference operator would become non-local because the
logarithmic function $\log w$ is not holomorphic on ${\cal D}_1$.
Hence, there is no nontrivial difference operator with the
requirements (i)$\sim$(iv).

Without the associative law, the no-go theorem can still
be proved as follows:

%
%
\begin{theorem}
\label{theorem2}
It is impossible to construct a lattice field theory in an infinite
lattice volume with a nontrivial product rule (\ref{product})
and a difference operator (\ref{diff}) that satisfy the
following three properties:
(i) translation invariance, (ii) locality and (iii) Leibniz rule.
\end{theorem}
%
%

Without referring to the associative law, it is impossible,
in general, to set $\hat{C}(v,w)=1$, as discussed in the
previous section. 
Instead, we use the existence of a domain
${\cal F}_2 = \{ (v,w) |\ \hat{C}(v,w) \ne 0\}$, on which
we have
%
\begin{equation}
\hat{D}(vw) - \hat{D}(v) - \hat{D}(w) = 0\, .
\label{eq27}
\end{equation}
%
The domain ${\cal F}_2$ should span 
a 2-dimensional complex domain in ${\cal D}_2$, 
otherwise $\hat{C}(v,w)$ would be
identically zero on ${\cal D}_2$.
The general solution to eq.(\ref{eq27}) is found 
as a logarithmic function on ${\cal F}_2$. 
The identity theorem enables to extend the domain  ${\cal F}_2$ 
to  ${\cal D}_2$.    
Thus, the difference operator cannot be holomorphic 
and hence is not local. 

We would like to make some comments here.
By remembering $w=e^{ip}$,  
the logarithmic function $\log w$ in (\ref{log-sol}) 
may be recognized as
a SLAC-type derivative \cite{SLAC} in infinite systems.
It is then interesting to note, as a corollary of our theorem, that
a SLAC-type derivative must be adopted 
as the difference operator if we construct a lattice field 
theory satisfying a Leibniz rule with a local product rule.
Although the proof is done in the case of an infinite 
lattice volume, the conclusion of the theorem can be kept 
even for general lattice theories with sufficiently large 
lattice size.   
We have proven the theorem for a one-dimensional theory and 
it is straightforward to generalize it for higher-dimensional 
cases.

\vspace{5mm}
%
%
%
\section{Multi-flavor extension}
%
%
%
We have presented a no-go theorem for general one-flavor systems.
It is not difficult to extend it to $N$-flavor systems.
A product rule and a difference operator are naturally
extended as
%
\begin{eqnarray}
(\phi \cdot \psi)_n^c 
  &\equiv& \sum_{l,m}\sum^{N}_{a,b=1} 
             C_{lmn}^{abc}\phi_{l}^{a} \psi_{m}^{b}\,  ,
\label{multi-product}\\
(D\phi)_n^b &\equiv& \sum_{m}\sum^{N}_{a=1} D_{mn}^{ab}\phi_{m}^{a}\, ,
\label{multi-diff}
\end{eqnarray}
%
where $a,b,c$ denote flavor indices.
A Leibniz rule in multi-flavor systems can be expressed as 
%
\begin{equation}
\sum_{k}\sum^{N}_{d=1}C^{abd}_{lmk}D^{dc}_{kn} =  
     \sum_{k}\sum^{N}_{d=1} C^{dbc}_{kmn}D^{ad}_{lk} + 
     \sum_{k}\sum^{N}_{d=1} C^{adc}_{lkn}D^{bd}_{mk}\, .
\label{multi-leibniz}
\end{equation}
%
The translation invariance and the locality 
for $C^{abc}_{lmn}$ and $D^{ab}_{mn}$ are defined in the same
way as in single flavor systems, and lead to the holomorphic
functions
%
\begin{eqnarray}
\hat{C}^{abc}(v,w) &\equiv&
  \sum_{l,m} C^{abc}(l,m) v^{l} w^{m},
  \label{multi-hol-C}\\
\hat{D}^{ab}(z) &\equiv&
  \sum_{m} D^{ab}(m) z^{m},
  \label{multi-hol-D}
\end{eqnarray}
%
where $C^{abc}_{lmn} = C^{abc}(l-n,m-n)$ and
$D^{ab}_{mn} = D^{ab}(m-n)$.
If the difference operator $D^{ab}_{mn}$ is independent of 
the flavor index, i.e.
%
\begin{equation}
D^{ab}_{mn} = \delta^{a,b}D_{mn}\, ,
\label{eq31}
\end{equation}
%
as in ordinary cases,
then our no-go theorem holds exactly as before.  
Even in more general case of $D^{ab}_{mn} = \delta^{a,b}D^{a}_{mn}$,
or equivalently  
%
\begin{equation}
\hat{D}^{ab}(z) = \delta^{a,b}\hat{D}^{a}(z)\, ,
\label{diago-D}
\end{equation}
%
our proof of the no-go theorem in the previous section
can be applicable. 
For $\hat{D}^{ab}(z)$ to act properly on $N$-flavor fields,
it is reasonable to assume that $\hat{D}^{ab}(z)$ can be
diagonalized with respect to the flavor indices, as in
eq.(\ref{diago-D}), by field redefinitions, which are
identical to similarity transformations on 
$\hat{D}^{ab}(z)$.
Thus, we succeeded in getting our no-go theorem for finite
flavor systems.

%
%
\begin{theorem}
\label{theorem3}
It is impossible to construct a lattice field theory 
of finite flavors in an infinite
lattice volume with a nontrivial product rule (\ref{multi-product})
and a difference operator (\ref{multi-diff}) that satisfy the
following three properties:
(i) translation invariance, (ii) locality and (iii) Leibniz rule.
\end{theorem}
%
%

Although a simple proof of the no-go theorem 3 was given above,
it will be worth presenting another proof of the theorem
that makes a connection clear between the product rule and
the difference operator through the Leibniz rule.
In terms of the holomorphic functions $\hat{C}^{abc}(v,w)$ and 
$\hat{D}^{ab}(z)$, the Leibniz rule
(\ref{multi-leibniz}) can be rewritten as
%
\begin{equation}
\sum^{N}_{d=1}\hat{C}^{abd}(v,w) \hat{D}^{dc}(vw) 
   = \sum^{N}_{d=1} \hat{C}^{dbc}(v,w) \hat{D}^{ad}(v) 
     + \sum^{N}_{d=1} \hat{C}^{adc}(v,w) \hat{D}^{bd}(w)\, .
\label{multi-hol-leibniz}
\end{equation}
%
It may be convenient, in the following analysis, to express
$\hat{C}^{abc}(v,w)$ and $\hat{D}^{ab}(z)$ into $N\times N$ matrix 
forms as
%
\begin{eqnarray}
\big( M^{b}(v,w) \big)_{ac} &\equiv& \hat{C}^{abc}(v,w)\,,
  \label{matrix-M}\\
\big( D(z) \big)_{ab} &\equiv& \hat{D}^{ab}(z)\,.
  \label{matrix-D}
\end{eqnarray}
%
In terms of them, the Leibniz rule (\ref{multi-hol-leibniz})
is further rewritten as
%
\begin{equation}
M^{b}(v,w) D(vw) = D(v) M^{b}(v,w)
     + \sum^{N}_{d=1} \hat{D}^{bd}(w) M^{d}(v,w)\, .
\label{matrix-leibniz}
\end{equation}
%
Setting $w=1$ in eq.(\ref{matrix-leibniz}) shows that
$D(v)$ commutes with $M^{b}(v,1)$, i.e.
%
\begin{equation}
[ M^{b}(v,1) , D(v) ] = 0 \qquad {\rm for}\ b=1,2,\cdots, N\,,
\label{[M,D]=0}
\end{equation}
%
where we have used
%
\begin{equation}
\hat{D}^{ab}(1) = 0\ \ {\rm or}\ \ D(1) = 0
\label{D(1)=0}
\end{equation}
%
which comes from the fact that $(D\phi)^{b}_{n}=0$ for any
constant field $\phi^{a}_{m}$.
Differentiating eq.(\ref{matrix-leibniz}) with respect
to $w$ and then setting $w=1$, we find
%
\begin{equation}
v\partial_{v}D(v) 
 = - \big[ M^{b}(v,1)^{-1}\partial_{w} M^{b}(v,w)\big|_{w=1},\ D(v) \big]
  + \sum^{N}_{d=1} \hat{D}^{\prime\,bd}(1)
     M^{b}(v,1)^{-1} M^{d}(v,1)\,,
\label{diff-eq-D}
\end{equation}
%
where
%
\begin{equation}
\hat{D}^{\prime\,bd}(1) \equiv \partial_{w}
  \hat{D}^{bd}(w) \big|_{w=1}\, .
\label{D'(1)}
\end{equation}
%
Here, we have used eqs.(\ref{[M,D]=0}) and (\ref{D(1)=0})
and assumed the existence of $M^{b}(v,1)^{-1}$.
We can regard eq.(\ref{diff-eq-D}) as a differential equation
for the difference operator $D(v)$ and formally solve it as
%
\begin{equation}
D(v) = \int^{v}_{1} \frac{dv'}{v'} U(v,v')\, X(v')\, U(v,v')^{-1},
\label{solution-D}
\end{equation}
%
where
%
\begin{eqnarray}
U(v,v') &\equiv& P \exp \bigg\{ - \int^{v}_{v'} \frac{dv''}{v''}
     M^{b}(v'',1)^{-1} \partial_{w} M^{b}(v'',w)
     \big|_{w=1}\ \bigg\}\, ,
  \label{U(v,v')}\\
X(v') &\equiv& \sum^{N}_{d=1} \hat{D}^{\prime\,bd}(1)
     M^{b}(v',1)^{-1} M^{d}(v',1)\,.
  \label{X(v)}
\end{eqnarray}
%
Here, $P$ denotes the path ordered product.

We should make several comments on eq.(\ref{solution-D}) here.
It is interesting to note that the relation (\ref{solution-D})
implies that the difference operator can completely be 
determined from information about the product rule (more
precisely, $M^{b}(v,1)$ and 
$\partial_{w}M^{b}(v,w)|_{w=1}$ ) and the
initial values $\hat{D}^{\prime\,bd}(1)$.
Although we have assumed the existence of $M^{b}(v,1)^{-1}$,
this will, however, be assured for a nontrivial product rule
because the function $\hat{C}^{abc}(v,w)$ (or $M^{b}(v,w)$)
are holomorphic and are not identically zero on ${\cal D}_{2}$.
We can always find a path of the integration 
in eq.(\ref{solution-D}) on which $M^{b}(v,1)^{-1}$ exists.
The third comment is that the right-hand-side of 
eq.(\ref{solution-D}) (or eq.(\ref{diff-eq-D})) depends 
on the flavor index $b$ (that is implicit in $U(v,v')$ and
$X(v')$), while the left-hand-side is independent of it.
This implies that the initial values $\hat{D}^{\prime\,bd}(1)$
are not freely chosen (see below). 
The final comment is that the expression (\ref{solution-D})
is not necessarily holomorphic and will not be, in particular,
single-valued.
This is indeed the case for a finite number of flavors as
we will see below.

In a naive continuum limit, $(D\phi)^{d}_{n}$ will be
expanded as follows:\footnote{
Eq.(\ref{continuumD}) will be obtained by noting that
$\phi^{b}_{m} \equiv \phi^{b}(ma) = e^{ma\partial_{x}} 
\phi^{b}(x)|_{x=0}$.
}
%
\begin{eqnarray}
\label{continuumD}
 \frac{1}{a}(D\phi)^{d}_{n} 
 \ \longrightarrow\ 
   \frac{1}{a}\sum_{b=1}^{N}\hat D^{bd}(1)\phi^{b}(x)
    + \sum_{b=1}^{N}\hat{D}^{\prime\,bd}(1)\partial_{x}\phi^{b}(x)
    + O\!\left(a\partial^{2}_{x}\phi(x)\right) ,
\end{eqnarray}
%
where $a$ is a lattice spacing.
We expect that the difference operator becomes a first-order
derivative in the naive continuum limit.
This requires, in addition to eq.(\ref{D(1)=0}),
that $\hat{D}^{\prime\,bd}(1)$ should be 
non-vanishing and possess $N$ independent eigenstates
with respect to the flavor indices, otherwise the
difference operator could not act on $N$-flavor fields
properly.
This observation guarantees that $\hat{D}^{\prime\,bd}(1)$
can be written in a diagonal form $\delta^{b,d}\beta^{b}$
by choosing an appropriate flavor basis.
Then, it follows from eq.(\ref{diff-eq-D}) at $v=1$ and
eq.(\ref{D(1)=0}) that 
$\beta^{b}$ should be flavor-independent, i.e. 
%
\begin{equation}
\hat{D}^{\prime\,bd}(1) = \delta^{b,d} \beta\,.
\label{D^{bd}(1)}
\end{equation}
%
Inserting eq.(\ref{D^{bd}(1)}) into eq.(\ref{solution-D}),
we finally find
%
\begin{equation}
\hat{D}^{ab}(v) = \delta^{a,b} \beta \log v\,.
\label{log-sol_2}
\end{equation}
%
This is a straightforward extension of the one-flavor case
(\ref{log-sol}).
Since the logarithmic function is not holomorphic 
on ${\cal D}_1$ and is not local, we thus arrive 
at the no-go theorem 3, as promised.

Before closing this section, it may be instructive to
discuss some properties of the product rule.
If the flavor indices and lattice sites of the product rule
are mutually independent, $M^{b}(v,w)$ are written
as $M^{b}(v,w) = M^{b}(1,1) F(v,w)$, where the complex
function $F(v,w)$ has no flavor index.\footnote{
If we further require the associative law, $F(v,w)$
is shown to be of the form
$F(v,w) = F(vw)/F(v)F(w)$ with a holomorphic
function $F(v)$.
This implies that $M^{b}(v,w)$ can reduce to $M^{b}(1,1)$
by local field redefinitions.
}
It immediately follows that the first term on the
right-hand-side of eq.(\ref{diff-eq-D}) vanishes and the
second term becomes independent of $v$.
We then find $\hat{D}^{ab}(v)$ to be proportional to
$\log v$, as found in eq.(\ref{log-sol_2}).
Thus we may conclude, from the observations in this section,
that at least a nontrivial connection between flavors and 
lattice sites for both of the product rule and the difference
operator is necessary in order to realize a Leibniz rule
to escape from our no-go theorem somehow
in a local lattice field theory.
This is the purpose of the next section.

\vspace{5mm}
%
%
%
\section{Matrix representation}
%
%
%
We have proved a no-go theorem about the existence of a 
product rule and a difference operator in any local lattice
field theories of finite flavors.
A way to escape from the no-go theorem is to consider infinite
flavor systems with a nontrivial connection between lattice
sites and flavors.
The proof given in the previous section cannot be applied
for an infinite number of flavors since the holomorphy/locality
of the product rule and the difference operator is not
necessarily preserved in diagonalizing $D^{\prime\,ab}(1)$
by field redefinitions.
This is because there is no guarantee that a linear
combination of an infinite number of holomorphic functions
in general is still holomorphic, 
though it is true for any linear combination of a finite 
number of holomorphic functions.
A nontrivial connection between lattice
sites and flavors is also inevitable because 
without any nontrivial relations the solution to 
eq.(\ref{diff-eq-D}) would be proportional to the 
non-holomorphic function $\log v$, 
irrespective of the size of flavors, 
as discussed in the previous section.

As a candidate, we present a representation of matrix fields
as lattice fields.
A matrix $\Phi_{ij}$ is identified with a lattice field 
$\phi^{a}_{n}$.
The correspondence of the indices are given by $a=i-j$ and 
$n=i+j$.
A product of two matrices, $(\Phi\Psi)_{ij}$, leads to a
product rule for lattice fields, $(\phi\cdot\psi)^{c}_{n}$, as
%
\begin{equation}
C^{abc}_{lmn} = \delta_{l-n,b} \delta_{n-m,a} \delta_{c,a+b}\,.
\label{matrix-product}
\end{equation}
%
It should be emphasized that the coefficients $C^{abc}_{lmn}$
give a nontrivial connection between lattice sites and flavor
indices and that the number of flavors has to be infinite
with an infinite lattice volume.
Since $C^{abc}_{lmn}$ are translationally invariant and local,
we can define holomorphic functions as 
%
\begin{equation}
\big( M^{b}(v,w)\big)_{ac}
 \equiv \hat{C}^{abc}(v,w) = \delta_{a+b,c} v^{b} w^{-a}.
\label{matrix-hol}
\end{equation}
%
It follows from eq.(\ref{solution-D}) that we find the expression
%
\begin{equation}
\hat{D}^{ac}(v) = \left\{
 \begin{array}{ll}
  \frac{\hat{D}^{\prime\, b,b-a+c}(1)}{2(a-c)}
   \big( v^{a-c} - v^{-a+c}\big),& \quad{\rm for}\ a \ne c\,,\\
  {\ }&{\ }\\
  \hat{D}^{\prime\, b,b}(1) \big( \log v + i2\pi \omega\big),&
   \quad{\rm for}\ a=c\,,
 \end{array}\right.
\label{matrix-sol-D}
\end{equation}
%
where $\omega$ is an integer which would come from the ambiguity
how to choose the path of the integration in eq.(\ref{solution-D}).
To avoid the logarithmic singularity, we have to impose the
conditions
%
\begin{equation}
\hat{D}^{\prime\, b,b}(1) = 0 \quad {\rm for}\ b=1,2,\cdots,N\,.
\label{restriction_D}
\end{equation}
%
We further require that $\hat{D}^{\prime\, ac}(1)$ depend
only on $a-c$.
This is because the right-hand-side of eq.(\ref{matrix-sol-D})
should be independent of the index $b$.
Thus, we have succeeded to obtain the difference operator
$\hat{D}^{ac}(v)$ to be of the form\footnote{
It will be instructive to note that $\hat{D}^{ac}(v)$ (or
$\hat{D}^{\prime\,ac}(1)$) given in eq.(\ref{matrix-hol-D}) 
could be diagonalized by field redefinitions 
only if we allow the product rule and the difference operator 
to be non-holomorphic/non-local.
}
%
\begin{equation}
\hat{D}^{ac}(v) = \tilde{d}(a-c)\, \big( v^{a-c} - v^{-a+c}\big)\,,
\label{matrix-hol-D}
\end{equation}
%
where 
$\tilde{d}(a-c) \equiv \hat{D}^{\prime\,b,b-a+c}(1) / 2(a-c)$.
Since $\hat{D}^{ac}(v)$ are holomorphic functions, we can have
a translationally invariant local difference operator 
$D^{ac}_{mn}$ satisfying the Leibniz rule as
%
\begin{equation}
D^{ac}_{mn} = \tilde{d}(a-c)\, \big( \delta_{m-n,a-c}
              - \delta_{m-n, -a+c}\big)\,.
\label{matrix-D}
\end{equation}
%
Although we can directly verify that the difference operator
(\ref{matrix-D}) and the product rule (\ref{matrix-product})
satisfy the Leibniz rule (\ref{multi-leibniz}) with infinite
flavors, it is more transparent in a matrix representation,
where the difference operator can be represented as the
commutator such that
%
\begin{equation}
(D\phi)^{b}_{n} \equiv [\, d, \Phi\, ]_{ij}
\label{matrix-commutator}
\end{equation}
%
with the identification of $b=i-j, n=i+j$ and
$d_{ij} = \tilde{d}(j-i)$.
Then, it is not difficult to see that the Leibniz rule
%
\begin{equation}
D(\phi\cdot\psi) = (D\phi)\cdot\psi + \phi\cdot(D\psi)
\label{matrix-Leibniz_1}
\end{equation}
%
is replaced by
%
\begin{equation}
[\,d, \Phi\Psi\,] = [\,d,\Phi\,]\Psi + \Phi[\,d, \Psi\,]\, ,
\label{matrix-Leibniz_2}
\end{equation}
%
which is rather a trivial relation.
In addition, we note that the product rule satisfies the
associative law
%
\begin{equation}
\sum_{j}\sum_{d} C^{abd}_{lmj}\,C^{dce}_{jnk} =
 \sum_{j}\sum_{d} C^{ade}_{ljk}\,C^{bcd}_{mnj}\, ,
 \label{matrix-associative}
\end{equation}
%
which is also trivially satisfied in the matrix representation,
as $\Phi(\Psi\Lambda) = (\Phi\Psi)\Lambda$.
It is interesting to note that this matrix representation
has already been applied to a quantum mechanical supersymmetric 
lattice model \cite{dublin}, where the lattice version of the full
supersymmetry is realized.

\vspace{5mm}
%
%
%
\section{Summary and discussions}
%
%
%
We have first shown the no-go theorem for general one-flavor
systems that it is impossible to construct a lattice theory
in an infinite lattice volume with a product rule and a
difference operator that satisfy the following three
properties: (i) translation invariance, (ii) locality and
(iii) Leibniz rule.
It turns out that the theorem holds even for multi-flavor
systems.
Our proof of the theorem shows that any difference operator 
satisfying the Leibniz rule can be determined from information
of the product rule and some initial data.
In fact, no difference operator is found to be local for
general multi-flavor systems.
If we allow the difference operator to be non-local, we can
construct it through the equation (\ref{solution-D}).

A breakthrough to evade the no-go theorem is to consider
an infinite number of flavors and a nontrivial connection
between lattice sites and flavors.
We presented such a lattice theory that infinite-flavor
fields are defined from matrix fields.
The product rule of two fields is just the product of 
two matrices and the difference operator that satisfies
the Leibniz rule is found to be a commutator with a 
matrix $d$.
They are all local and translationally invariant with
respect to lattice sites.
Furthermore, the product rule turns out to satisfy the 
associative law.
In our forthcoming papers, we shall discuss the lattice
theory equipped with the above tools and clarify their 
properties, in detail.
In particular, we shall present lattice supersymmetric 
models which realize the lattice version of the full
supersymmetry.

Another way we could incorporate an infinite number of 
flavors is to consider extra dimensions where locality 
or translational invariance in the whole space would 
be somehow broken whereas those of the target
space should be preserved. 
We also hope to report some attempts in this direction 
elsewhere.

Other possibilities to escape from the no-go theorem may
be to generalize translational invariance and locality.
A candidate is a lattice formulation based on non-commutative
geometry \cite{conne} .
Further analysis in translational invariance and locality
should be done.

\vspace{10mm}
%
%
%
%
%
%
\acknowledgments
One (H.S.) of the authors thanks to Dr. N. Ukita for
discussions at an earlier stage. 
This work is supported in part by the Grant-in-Aid for Scientific 
Research (No.19540272 (M.K.), No.18540275 (M.S.) and No.17540242 (H.S.))
by the Japanese Ministry of Education, Science, Sports and Culture.

\vspace{10mm}
%
%
%
\appendix
\section{Factorization of the product rule}
%
%
%
In this appendix, we prove the relation (\ref{sol-assoc}).

Setting $v=1$ in eq.(\ref{hol-associative-law}) leads to the
relation $\hat{C}(1,w) \hat{C}(w,z) = \hat{C}(1,wz) \hat{C}(w,z)$,
which reduces to $\hat{C}(1,w) = \hat{C}(1,wz)$ on the 
2-dimensional complex domain 
${\cal F}_{2} = \{(w,z)|\,\hat{C}(w,z) \ne 0\,\}$.
This implies that $\hat{C}(1,w)$ is independent of $w$ on
${\cal D}_1$. 
Setting $v=1/w$ in eq.(\ref{hol-associative-law}) leads to the
relation $\hat{C}(1/w,wz) = \hat{C}(1/w,w) \hat{C}(1,z)/
\hat{C}(w,z)$ on ${\cal F}_2$.
If $\hat{C}(1,z)$ is zero, then $\hat{C}(1/w,wz)$ would be
identically zero.
However, this cannot be the case for a nontrivial product
rule $\hat{C}(v,w)$.
Therefore, $\hat{C}(1,z)$ cannot be zero.
This argument is applicable for $\hat{C}(v,1)$.
We thus conclude
%
\begin{equation}
\hat{C}(1,z) = \hat{C}(v,1) = \alpha\,,
 \label{eqA1}
\end{equation}
%
where $\alpha$ is a nonzero constant.

By differentiating eq.(\ref{hol-associative-law}) with 
respect to $v$ and then taking $v=1$, we have
%
\begin{equation}
f(w) \hat{C}(w,z) + \alpha w \partial_{w}\hat{C}(w,z)
 = f(wz) \hat{C}(w,z)\,,
 \label{eqA2}
\end{equation}
%
where $f(w) \equiv \partial_{v} \hat{C}(v,w)|_{v=1}$.
We regard eq.(\ref{eqA2}) as a differential equation for
$\hat{C}(w,z)$ with a given function $f(w)$.
With the initial condition (\ref{eqA1}), the differential 
equation (\ref{eqA2}) can easily be solved as
%
\begin{equation}
\hat{C}(w,z) = \alpha\exp\bigg\{
 \frac{1}{\alpha} \int^{w}_{1} \frac{du}{u}
   \Big( f(uz) - f(u) \Big) \bigg\}\,.
 \label{eqA3}
\end{equation}
%
We should emphasize that the expression (\ref{eqA3}) is
well-defined and that there is no ambiguity in the definition
of the integral with respect to $u$ because
%
\begin{equation}
\oint \frac{du}{u} \Big( f(uz) - f(u) \Big) = 0
 \label{eqA4}
\end{equation}
%
for any closed loop on ${\cal D}_1$.

Since $f(u)$ is holomorphic on ${\cal D}_1$, it can be
uniquely expanded in the Laurent series as
%
\begin{equation}
f(u) = \sum^{\infty}_{n=-\infty} f_{n}\, u^{n}.
 \label{eqA5}
\end{equation}
%
Let us introduce a new holomorphic function $\tilde{f}(u)$ as
%
\begin{equation}
\tilde{f}(u) \equiv f(u) - f_{0}\,.
 \label{eqA6}
\end{equation}
%
Then, for any closed loop on ${\cal D}_1$ we find
%
\begin{equation}
\oint \frac{du}{u} \tilde{f}(u) 
  = \oint \frac{du}{u} \sum_{n\ne 0} f_{n}\,u^{n} = 0\,.
 \label{eqA7}
\end{equation}
%
In terms of $\tilde{f}(u)$, eq.(\ref{eqA3}) can be rewritten as
%
\begin{eqnarray}
\hat{C}(w,z) 
  &=& \alpha \exp\bigg\{ \frac{1}{\alpha}
        \int^{w}_{1} \frac{du}{u}\Big(f(uz) - f(u)\Big)\bigg\}
        \nonumber\\
  &=& \alpha \exp\bigg\{ \frac{1}{\alpha}
        \int^{w}_{1} \frac{du}{u}\Big(
        \tilde{f}(uz) - \tilde{f}(u)\Big)\bigg\}
        \nonumber\\
  &=& \frac{
            \frac{1}{\alpha} \exp\bigg\{ \frac{1}{\alpha}
            \int^{wz}_{1} \frac{du}{u}\tilde{f}(u)\bigg\}
           }
           {
            \frac{1}{\alpha} \exp\bigg\{ \frac{1}{\alpha}
            \int^{w}_{1} \frac{du}{u}\tilde{f}(u)\bigg\}\,
            \frac{1}{\alpha} \exp\bigg\{ \frac{1}{\alpha}
            \int^{z}_{1} \frac{du}{u}\tilde{f}(u)\bigg\}
           }  
        \nonumber\\    
  &\equiv& \frac{F(wz)}{F(w) F(z)}\,.                 
\label{eqA8}
\end{eqnarray}
%
We note that the function $F(w)$ is well-defined and that
there is no ambiguity for the integral because of the
property (\ref{eqA7}).

\vspace{10mm}
%
%
%
%
%
%

%
%
%

%
%
%

\begin{thebibliography}{99}
\bibitem{Feo2003} A. Feo,
\emph{Supersymmetry on the lattice},
\npps{119}{2003}{198}
[\heplat{0210015}]. 
%
\bibitem{Kaplan2004} D.B. Kaplan,
\emph{Recent developments in lattice supersymmetry},
\npps{129}{2004}{109}
[\heplat{0309099}];
\emph{Supersymmetry on the lattice},
\emph{Eur. Phys. J. ST.} \textbf{152} (2007) 89.
%
\bibitem{Giedt} J. Giedt,
\emph{Advances and applications of lattice supersymmetry},
\emph{PoS LAT2006} (2006) 008
[\heplat{0701006}]. 
%
\bibitem{Dondi-Nicolai}  P.H. Dondi and H. Nicolai, 
\emph{Lattice Supersymmetry},
\nc{A41}{1977}{1}. 
%
\bibitem{Fujikawa}  K. Fujikawa, 
\emph{Supersymmetry on the lattice and the Leibniz rule},
\npb{636}{2002}{80}
[\hepth{0205095}];
\emph{$N = 2$ Wess-Zumino model on the $d = 2$ Euclidean lattice},
\prd{66}{2002}{074510}
[\heplat{0208015}].
%
\bibitem{Kawamoto2005}  A. d'Adda, I. Kanamori, N. Kawamoto, 
and K. Nagata,
\emph{Twisted superspace on a lattice},
\npb{707}{2005}{100}
[\heplat{0406029}]. 
%
\bibitem{Kawamoto2006} A. d'Adda, I. Kanamori, N. Kawamoto, 
and K. Nagata,
\emph{Exact extended supersymmetry on a lattice:
Twisted N=2 super Yang-Mills in two dimensions},
\plb{633}{2006}{645}
[\heplat{0507029}]. 
%
\bibitem{Bruckmann2006}  F. Bruckmann and M. de Koh, 
\emph{Noncommutativity approach to supersymmetry on the lattice: 
SUSY quantum mechanics and an inconsistency},
\prd{73}{2006}{074511}
[\heplat{0603003}]. 
%
\bibitem{Bruckmann2007}  F. Bruckmann, S. Catterall and M. de Koh, 
\emph{Critique of the link approach to exact lattice supersymmetry},
\prd{75}{2007}{045016}
[\heplat{0611001}]. 
%
\bibitem{SLAC} S.D. Drell, M. Weinstein and S. Yankielowicz, 
\emph{Strong-coupling field theories. II. 
Fermions and gauge fields on a lattice}, 
\prd{14}{1976}{1627}.
%
\bibitem{Bartels-Bronzan}  J. Bartels and J.B. Bronzan, 
\emph{Supersymmetry on a lattice},
\prd{28}{1983}{818}. 
%
\bibitem{Nojiri}  S. Nojiri, 
\emph{Continuous \lq Translation\rq\ and Supersymmetry on the
lattice},
\ptp{74}{1985}{819};\  
\emph{The Spontaneous Breaking of Supersymmetry on the
Finite Lattice},
\ptp{74}{1985}{1124};\  
%
\bibitem{Bouguenaya-Fairlie}  Y. Bouguenaya and D.B. Fairlie, 
\emph{A finite difference scheme with a Leibniz rule},
\jpha{19}{1986}{1049}. 
%
\bibitem{So-Ukita} H. So and N. Ukita, a talk in a workshop on 
\lq\lq Supersymmetry, 
Chiral Symmetry and Lattice Gauge Theory" (SUCAL 99),
Niigata, Japan, 17-19, Nov. 1999.  
%
\bibitem{Ginsparg-Wilson} P.H. Ginsparg and K.G. Wilson, 
\emph{A Remnant of Chiral Symmetry on the Lattice}, 
\prd{25}{1982}{2649}.
%
\bibitem{Neuberger} H. Neuberger, 
\emph{Exactly massless quarks on the lattice}, 
\plb{417}{1998}{141}
[\heplat{9707022}];
\emph{More about exactly massless quarks on the lattice}, 
\plb{427}{1998}{353}
[\heplat{9801031}].
%
\bibitem{HJL} P. Hernandez, K. Jansen and M. Luscher, 
\emph{Locality properties of Neuberger's lattice Dirac operator}, 
\npb{552}{1999}{363}
[\heplat{9808010}].
%
\bibitem{dublin} M. Kato, M. Sakamoto and H. So, 
\emph{Leibniz rule and exact supersymmetry on lattice: 
A Case of supersymmetrical quantum mechanics}, 
\emph{PoS LAT2005} (2006) 274
[\heplat{0509149}].
%
\bibitem{conne} A. Connes,  
\emph{Noncommutative Geometry}, Academic Press, 1994.\\
See also:
M.~Dubois-Violette, R.~Kerner and J.~Madore,
\emph{Noncommutative Differential Geometry Of Matrix Algebras},
\jmp{31}{1990}{316}.\\
A.~Dimakis, F.~Mueller-Hoissen and T.~Striker,
\emph{Noncommutative Differential Calculus And Lattice Gauge Theory},
\jpha{26}{1993}{1927}.
%
\end{thebibliography}
\end{document}